\documentclass[showpreprint,preprint,showpacs,preprintnumbers,amsmath,amssymb]{revtex4}

\usepackage{amsmath} 
\usepackage{amssymb}
\newcommand{\ud}{\mathrm{d}}
\usepackage{epsf}
\usepackage{epsfig}
\usepackage{graphicx}
\usepackage{dcolumn}
\usepackage{bm}
\usepackage{psfrag}

\begin{document} 
\preprint{\emph{
This article is identical to the version
that appeared in Appl. Phys. Lett. {\bf 88}, 203506 (2006) .}
}
\preprint{
\emph{The published version can be found at http://apl.aip.org/.}
}
\draft
\title{Radio-frequency point-contact electrometer}

\draft
\author{Hua Qin}\email{qin1@wisc.edu}
\altaffiliation[Current address: ]{Laboratory for Molecular-Scale Engineering, 
Department of Electrical and Computer Engineering, University of Wisconsin-Madison, 
1415 Engineering Drive, Madison, WI-53706, USA}
\affiliation{Microelectronics Research Centre, Cavendish Laboratory, 
Department of Physics, University of Cambridge, 
Madingley Road, Cambridge CB3 0HE, United Kingdom}
\author{David A. Williams} 
\affiliation{Hitachi Cambridge Laboratory, Cavendish Laboratory, 
Madingley Road, Cambridge CB3 0HE, United Kingdom}

\date{\today} 

\begin{abstract} 
We fabricate and characterize a radio-frequency semiconductor 
point-contact electrometer (RF-PC) analogous to radio-frequency 
single-electron transistors [RF-SETs, see Science {\bf 280}, 1238 
(1998)]. The point contact is formed by surface Schottky gates 
in a two-dimensional electron gas (2DEG) in an AlGaAs/GaAs heterostructure. 
In the present setup, the PC is operating as a simple voltage-controlled 
resistor rather than a quantum point contact (QPC) and demonstrates 
a charge-sensitivity about $2\times 10^{-1}~\mathrm{e/\sqrt{Hz}}$ 
at a bandwidth of $30~\mathrm{kHz}$ without the use of a cryogenic RF preamplifier. 
Since the impedance of a typical point-contact 
device is much lower than the impedance of the typical SET, 
a semiconductor-based RF-PC, equipped with practical cryogenic RF preamplifiers, 
could realize an ultra-fast and ultra-sensitive electrometer. 
\end{abstract}


\maketitle

A fast and sensitive electrometer is desirable for sensor applications, 
for studying charge dynamics in mesoscopic electron devices and for read-out 
devices in quantum-information processing. The radio-frequency 
single-electron transistor (RF-SET), invented by Schoelkopf {\it et al} 
in 1998, is the most sensitive {\it and} fastest electrometer known 
in the literature\cite{schoelkopf, rimberg, buehler}. 
The power of such an RF-SET comes from 
the integration of an aluminum SET, with ultra-high charge sensitivity, 
into an RF transmission line, with an embedded LC resonant circuit~\cite{schoelkopf}. 
Similar ideas have been widely adopted in many different applications, 
e.g., to sense the single-electron tunneling in a single-wall carbon 
nanotube~\cite{biercuk} or to count moving particles 
in a micro-fluidic channel~\cite{wood}. 
So far, RF-SETs are commonly 
made of aluminum and fabricated close to the devices under test. 
RF-SET electrometers made of host semiconductor materials will 
significantly simplify the fabrication procedure 
and the architecture of integrated device-electrometer system. 
However, a practical LC-resonant circuit embedded 
in a $50~\mathrm{\Omega}$ transmission line requires 
the SET impedance to be low enough ($R < 100 ~\mathrm{k\Omega}$) 
to maintain a high sensitivity (see below). 
Although there is one successful semiconductor-based RF-SET reported~\cite{fujisawa}, 
it turns out very challenging to build an RF-SET from resistive 
semiconductor SETs ($R \sim 500~\mathrm{k\Omega}$) than 
to make one from metallic counterparts ($R \sim 50~\mathrm{k\Omega}$).

Unlike SETs, which require large tunnel barrier impedances in 
order to localize the electron on the island, a point contact (PC) 
operates as a conventional field-effect transistor but with 
very low gate-channel capacitance and very low channel charge. 
This enables the PC to use a much lower channel resistance than 
is possible with SETs and so is ideal for integrating into a 
resonant RF transmission line (forming an RF-PC analogous to RF-SETs). 
Under appropriate conditions, the conductance of a PC becomes quantized: 
$G_N=2e^2N/h = N \times (12.9~\mathrm{k\Omega})^{-1}$, 
$N=1, 2, ...$, and the device is called quantum point contact (QPC). 
The sharp transition regions between neighboring conductance plateaus 
allow even greater sensitivity in the detection of 
charge fluctuations in nearby electron devices to be realized~\cite{field, elzerman}. 
Recent results reported in Ref.~\cite{vandersypen} have shown real-time 
observations of single-electron tunneling events in a quantum 
dot detected by a capacitively coupled QPC electrometer. 
In that experiment, a more direct approach was applied: 
a room temperature $I-V$ converter was applied to monitor 
the electrical current through the QPC. 
Here we demonstrate the first implementation of an RF-PC electrometer 
by integrating a PC device with a resonant RF transmission line. 
The electrometer was characterized at both dc and RF limit. 


Fig.~\ref{fig:1}(a) shows the circuit diagram of an RF-PC.   
Two inductors and a PC device are embedded in the RF transmission line.  
The inductors ($L$) and the capacitance ($C$) of the PC form an LC resonant circuit. 
The LC circuit introduces a discontinuity in the electromagnetic 
impedance in the transmission line (with a characteristic impedance $z_0$), 
which reflects an incident RF signal. 
The reflection coefficient is $\Gamma = (z-z_0)/(z+z_0)$, 
where $z$ is the input impedance: $z = z_1 +z_2(z_0+z_1)/(z_0+z_1+z_2)$ 
with $z_1 = j2\pi f L$, $z_2 = R_q/(1+j2\pi f R_qC)$, 
and $R_q=R_q(V_g)$ being the resistance of the PC, which is controlled by the gate voltage $V_g$. 
Neglecting any phase shift, the transmitted signal is given by $v_o = T(f,R_q) v_i\cos(2\pi ft)$, 
where  $v_i\cos(2\pi ft)$ is the input RF signal 
and $T(f,R_q)$ is the transmission coefficient: 
$T(f,R_q) =  \big \{ (1-|\Gamma|^2)/[1+z_0/R_q+(2\pi f)^2L^2/R_qz_0] \big \}^{1/2}$.
This circuit has a resonance at $f_0 \approx (2/LC)^{1/2}/2\pi$ 
with a $Q$-factor of $Q=2\pi f_0L/z_0$. 
Since the transmission coefficient at resonance, $T(f_0,R_q)$,  
is controlled by the gate voltage $V_g$,  
a small ac signal $v_g\cos(2\pi \xi t)$, 
superimposed on a dc working-point gate voltage $V_g^0$, 
will generate two sidebands (at $f_0 \pm \xi$) in the output spectrum:  
$v_{os} = \frac{v_iv_g}{4} 
\big\{\frac{\ud T(f_0,R_q)}{\ud R_q} |\frac{\ud R_q}{\ud V_g}|\big\}_{V_g=V_g^0} 
\cos\big[2\pi(f_0 \pm \xi)t\big ]$. 
The sensitivity of the output signal to small changes in the gate voltage 
is proportional to both 
$\ud T/\ud R_q$ and $|\ud R_q/\ud V_g|$. 
The first term is determined by the characteristics of the RF transmission line,   
while the second term is determined by the characteristic of the PC. 

In our experiment, the RF transmission line is realized 
as a coplanar waveguide ($z_0=50~\mathrm{\Omega}$)
on a high-frequency printed-circuit board (PCB), 
as shown in Fig.~\ref{fig:1}(b). Two high-Q on-chip inductors 
($L=100~\mathrm{nH}$) are inserted into the coplanar waveguide 
with small spacing. The semiconductor substrate containing the 
PC device is mounted and electrically connected to the PCB next 
to the junction of the two inductors, circuit using short gold 
bond wires. The PC device was fabricated using a two-dimensional 
electron gas (2DEG) AlGaAs/GaAs heterostructure, 
with an electron density of $1.7\times 10^{15}~\mathrm{m^{-2}}$ about $90~\mathrm{nm}$  
beneath the surface and a mobility of $80~\mathrm{m^{2}/Vs}$ at $4.2~\mathrm{K}$. 
The PC is defined by two reverse-biased Schottky 
gates connected to the coplanar waveguide, as shown in Fig.~\ref{fig:1}. 
The entire circuit is enclosed in an oxygen-free-high-conductivity copper box, 
with four SMA connectors. 
The input RF signal was supplied by either an HP 4396A network/spectrum analyzer 
or an HP 83711A synthesized continuous-wave generator. 
The input power was fixed at $-40~\mathrm{dBm}$, 
which corresponds to an ac voltage of $v_i=3.2~\mathrm{mV}$. 
The transmitted RF signal was monitored by the spectrum analyzer. 
No net drain-source bias was applied in this experiment ($V_{ds}=0$). 
However, a $17~\mathrm{Hz}$ sinusoidal signal 
with the root-mean-square amplitude of $158~\mathrm{\mu V}$ 
was superimposed onto the input RF signal using a bias tee. This 
allowed of the measurement of PC's differential 
conductance ($G=\ud I_{ds}/\ud V_{ds} \approx 1/R_q$) 
by an ITHACO 1211 current preamplifier together with an EG\&G 5210 lock-in amplifier. 
All measurements were performed at $4.2~\mathrm{K}$.

The transmission coefficient as a function of the frequency 
measured by the network analyzer is shown in Fig.~\ref{fig:1}(c).  
The resonance at $f_0=810~\mathrm{MHz}$ suggests 
a total capacitance of $C \approx 0.786~\mathrm{pF}$~\cite{cap}.  
The resonance has a $Q$-factor about 10.  
The dashed line in Fig.~\ref{fig:1}(c) is a simulation 
based on the circuit model shown in Fig.~\ref{fig:1}(a) given $R_q = 100~\mathrm{M\Omega}$. 
In addition to the attenuation in the point contact, 
there is also an attenuation of about $6~\mathrm{dB}$ 
in the coaxial cables, the coplanar waveguide and the inductors. 
  

By tuning the dc gate voltage, the PC's resistance and 
hence the amplitude of the transmission are varied, 
as shown in Fig.~\ref{fig:1}(c). 
A detailed correspondence between the resistance  
and the resonant transmission coefficient [$T(f_0)$] 
is shown in Fig.~\ref{fig:2}(a).  
The resistance increases with  
a more negative gate voltage and 
finally reaches the pinch-off state at around $-1.65~\mathrm{V}$.
Accordingly, the transmission is enhanced by the increase of resistance.
However, the measured transmission does not saturate beyond the pinch-off voltage and 
is lower than the calculated values based on the lumped-circuit model shown in Fig.~\ref{fig:1}(a). 
This discrepancy is caused by the assumptions of this simple circuit model:  
The resistance measured at quasi-dc limit and a shunted capacitance are 
not sufficient for modeling the RF characteristics. 
A network of stray capacitors and resistors could be a better model; 
nevertheless the present model fits very well to the data 
for $R_q < 23~\mathrm{k\Omega}$ (i.e., $V_g > -1.3~\mathrm{V}$) 
by taking into account an extra attenuation of $7~\mathrm{dB}$ 
resulting from the stray network. 
With higher resistance, the transmission becomes more sensitive to variations in capacitance. 
A careful analysis of the resonant frequency reveals 
that a shift from $786~\mathrm{MHz}$ to $810~\mathrm{MHz}$ occurs 
when $V_g$ is varied from $0~\mathrm{V}$ to $-2.5~\mathrm{V}$. 
The total capacitance calculated from this frequency shift is shown in Fig.~\ref{fig:2}(b). 
The plateau around $-1.5~\mathrm{V}$ 
indicates complete depletion of 2DEG immediately under the gates. 

In order to examine the sensitivity of this RF-PC electrometer, 
a sinusoidal signal at a frequency of $\xi =1~\mathrm{MHz}$ and 
a power of $-6.0~\mathrm{dBm}$ was superimposed on a dc gate 
voltage of $V_g^0 = -1.405~\mathrm{V}$. 
The input RF signal was fixed at $f_0=813~\mathrm{MHz}$. 
In Fig.~\ref{fig:3}(a), two sidebands at $813 \pm 1~\mathrm{MHz}$ are shown in the output spectrum 
measured by the spectrum analyzer with a bandwidth of $30~\mathrm{kHz}$.
The sidebands come from the modulation of the PC's 
resistance ($R_q^0 \approx 42.4~\mathrm{k\Omega}$) by the ac gate voltage. 
Fig.~\ref{fig:3}(b) shows the measured and 
calculated sideband amplitudes, which decrease as the PC's resistance becomes larger.  
The oscillatory feature shown in the measured data is 
a result of the fluctuation in effective ac gate voltage 
caused by the changing stray capacitance as discussed earlier.  
As shown in Fig.~\ref{fig:3}(a), the noise-floor is about $-102~\mathrm{dBm}$, 
corresponding to a noise voltage of $1.8~\mathrm{\mu V}$. 
The sidebands have an amplitude of about $-90~\mathrm{dBm}$, 
i.e., the signal voltage is about $7~\mathrm{\mu V}$ and 
the signal-to-noise ratio (SNR) is $12~\mathrm{dB}$.  
The ac gate voltage can be estimated from the amplitude of the sidebands 
to be about $v_g \approx 63~\mathrm{mV}$. 
As an order-of-magnitude estimate, such an ac signal 
induces a charge variation of about $\Delta q = 0.707C_gv_g \approx 140\mathrm{e}$, 
assuming a gate capacitance of $C_g=500~\mathrm{aF}$~\cite{qin}. 
The charge sensitivity of this RF-PC at $V_g^0$ can be estimated as    
$\delta q = \Delta q/(\sqrt{B}{10}^{SNR/20}) \approx 2 \times 10^{-1}\mathrm{e}/\sqrt{\mathrm{Hz}}$ 
at $B=30~\mathrm{kHz}$. 


The present sensitivity is limited by the noise from within the 
spectrum analyzer, which is much larger than the intrinsic noises generated 
by the PC itself, such as shot noise and thermal noise~\cite{noise}. 
The sensitivity can be significantly increased by using 
a cryogenic RF preamplifier, which has a low noise figure.  
Based on the present setup, an additional cold RF preamplifier 
with a noise temperature of $20~\mathrm{K}$ and a gain of $20~\mathrm{dB}$ 
will allow us to reduce the input RF signal 
from $-40~\mathrm{dBm}$ to $-60~\mathrm{dBm}$. 
The shot noise is thus reduced by a factor of ten and 
the thermal noise at $4.2~\mathrm{K}$ becomes the limiting factor. 
The overall charge sensitivity will be increased by a factor of 25.  
Furthermore, by carefully engineering the PC, 
the intrinsic sensitivity ($|\ud R_q/\ud V_g|$) could 
be increased by a factor of ten~\cite{elzerman}. 
So, an overall charge sensitivity of 
$1 \times 10^{-3}~\mathrm{e/\sqrt{Hz}}$ at $30~\mathrm{kHz}$ could be expected.

With a fixed design of the LC circuit and a given cryogenic RF preamplifier, 
the main way of improving the sensitivity is to engineer the PC  
so that both a larger ($|\ud R_q/\ud V_g|$) and a smaller $R_q$ 
could be achieved at the chosen working point $V_g^0$. 
As shown in Fig.~\ref{fig:3}(c), 
the benefit of a smaller resistance is a larger $\ud T/\ud R_q$. 
Furthermore, a smaller resistance means less thermal and shot noise. 
To have a large ($|\ud R_q/\ud V_g|$), one can set the working-point gate voltage 
in the middle of the sharp transition region between two quantized conductance plateaus. 
However, in this case, the dynamic range of the electrometer is strongly limited by 
the narrow transition range. 
A larger ($|\ud R_q/\ud V_g|$) can also be achieved 
by carefully engineering the design of PC even without conductance quantization. 
In this case, the RF-PC could provide both a high sensitivity and a large dynamic range. 
Comparing to resistive RF-SETs, an RF-PC/RF-QPC made of the host semiconductor material 
is much easier to fabricate and integrate into the system under test. 
Furthermore, the quantum electron-waveguide nature of a QPC allows  
continuous weak quantum measurements~\cite{korotkov, jordan}. 
In addition to SETs and PCs, any controllable resistor/capacitor/inductor device 
could be integrated into this RF circuit as a fast and sensitive sensor. 
Some of the possible devices are MOSFET transistors, 
Zener diodes, and microelectromechanical capacitor sensors. 

In conclusion, we have constructed and characterized an RF-PC 
with a charge sensitivity of $2 \times 10^{-1}~\mathrm{e/\sqrt{Hz}}$ 
at $30~\mathrm{kHz}$ which is limited by the noise from the RF 
measurement instrument at room temperature. 
By introducing a cryogenic RF preamplifier, lowering the input RF power, 
and improving the point-contact to have steeper pinch-off characteristic, 
the charge sensitivity could be increased to $1 \times 10^{-3}~\mathrm{e/\sqrt{Hz}}$ or even higher. 
In addition to the fact that QPC electrometers promise an ultra-high sensitivity at quantum limit,  
we emphasize that it is easier to build an RF-PC than conventional RF-SETs. 
We also point out that any device, such as a variable resistor with a sharp control slope 
and a resistance below $100~\mathrm{k\Omega}$, 
could be easily integrated into a similar RF circuit 
to form a both fast and sensitive detector/sensor. 

This work is supported by the U.K. Department of Trade and Industry 
and Hitachi Europe Limited under the Foresight LINK 
project ``Nanoelectronics at the Quantum Edge''. 
The authors thank Dr. David G. Hasko for stimulating discussions 
and critical proof reading of the manuscript, 
and thank Dr. Richard J. Collier for the discussions on RF measurements.

\newpage
\begin{figure}
\caption{
(a) Circuit diagram for the RF-PC electrometer and a scanning-electron micrograph of the PC.  
(b) Realization of the RF-PC on a high-frequency printed-circuit board. 
(c) Measured transmission coefficient 
as a function of the frequency at $4.2~\mathrm{K}$ (solid curves). 
From bottom to top, the dc gate voltage is $0~\mathrm{V}$, $-1.0~\mathrm{V}$, 
$-1.5~\mathrm{V}$, and $-2.5~\mathrm{V}$, respectively.   
Calculated transmission coefficient is shown as the dashed curve.
} \label{fig:1}

\caption{
(a) The correspondence between 
the resonant transmission coefficient and the PC's resistance. 
The vertical dotted line marks the pinch-off voltage ($-1.65~\mathrm{V}$). 
The solid curve represents the calculated resonant transmission amplitude. 
(b) The total capacitance of the PC is varied by the gate voltage. 
} \label{fig:2}

\caption{
(a) Transmission spectrum measured by the spectrum analyzer. 
(b) Linear-scale amplitude of the sidebands as a function of the gate voltage
compared with the simulations (solid curves). 
For clarity, the data for sideband at $814~\mathrm{MHz}$ are shifted upward by $20~\mathrm{\mu V}$. 
(c) Measured $\ud T/\ud R_q - R_q$ curve is compared with the simulation (solid curve). 
The dashed lines are calculated shot noise and thermal noise across the PC. 
The PC has a resistance of $42.4~\mathrm{k\Omega}$ and 
is biased with a constant voltage $v_{ds}^{rms} \approx 17~\mathrm{mV}$ at $4.2~\mathrm{K}$. 
} \label{fig:3}

\end{figure}

\begin{figure}[!p] 
\vspace{2 cm}
\includegraphics[width=.8\textwidth]{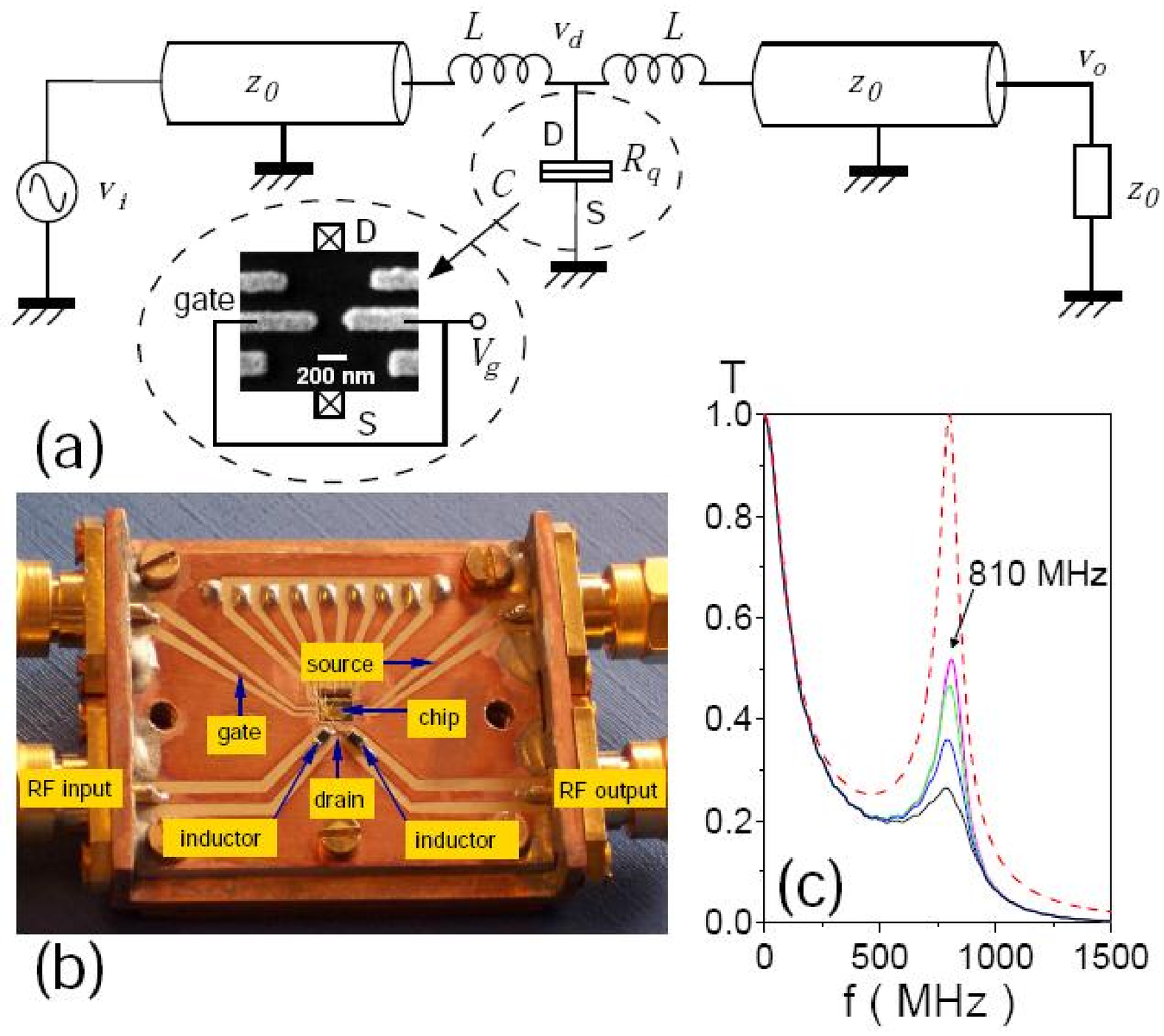}
\center{Qin {\it et al:~} Figure 1/3}
\end{figure}

\begin{figure}[!p] 
\vspace{4 cm}
\includegraphics[width=.8\textwidth]{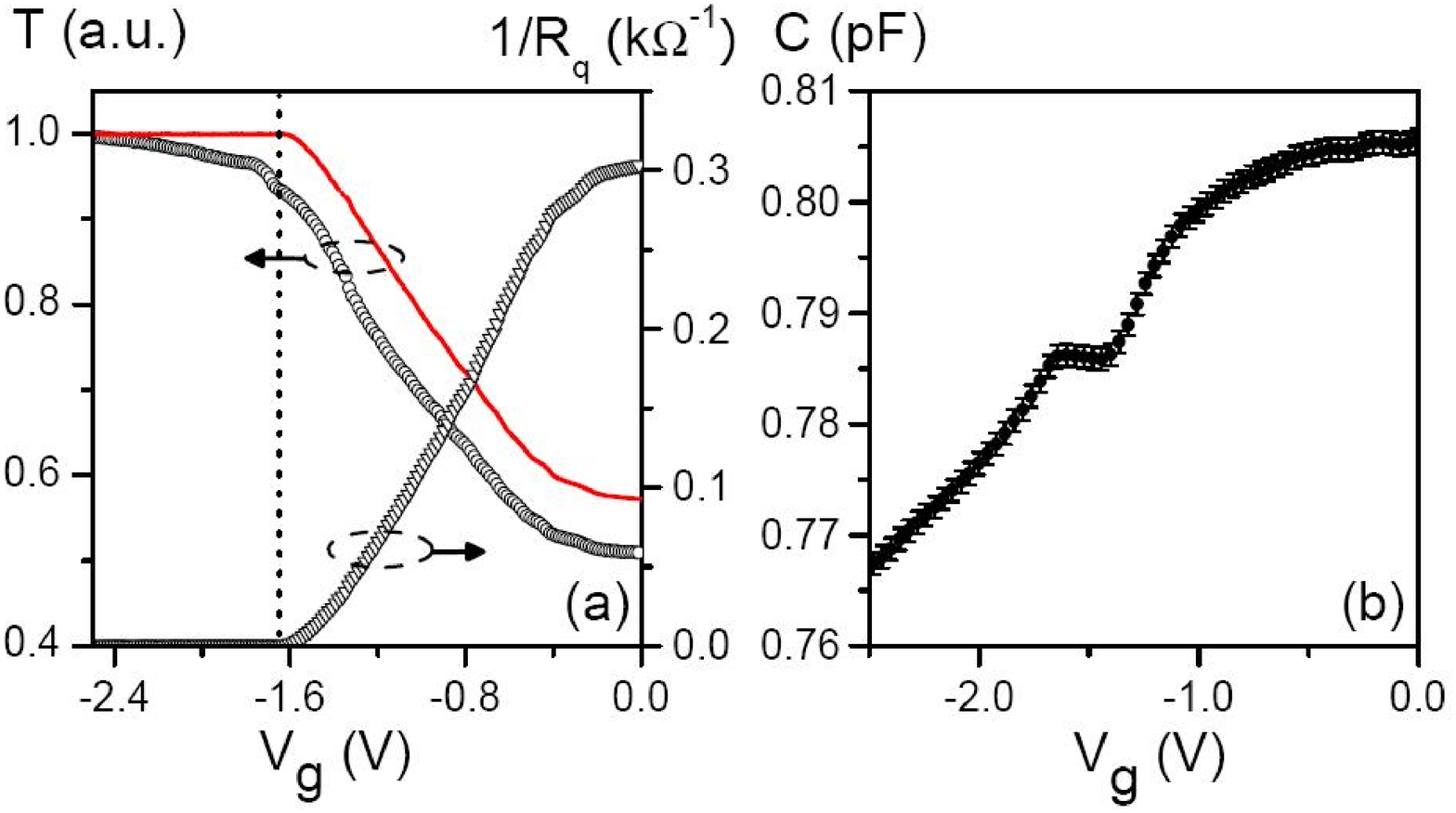}
\center{Qin {\it et al:~} Figure 2/3}
\end{figure}

\begin{figure}[!p] 
\vspace{4 cm}
\includegraphics[width=.8\textwidth]{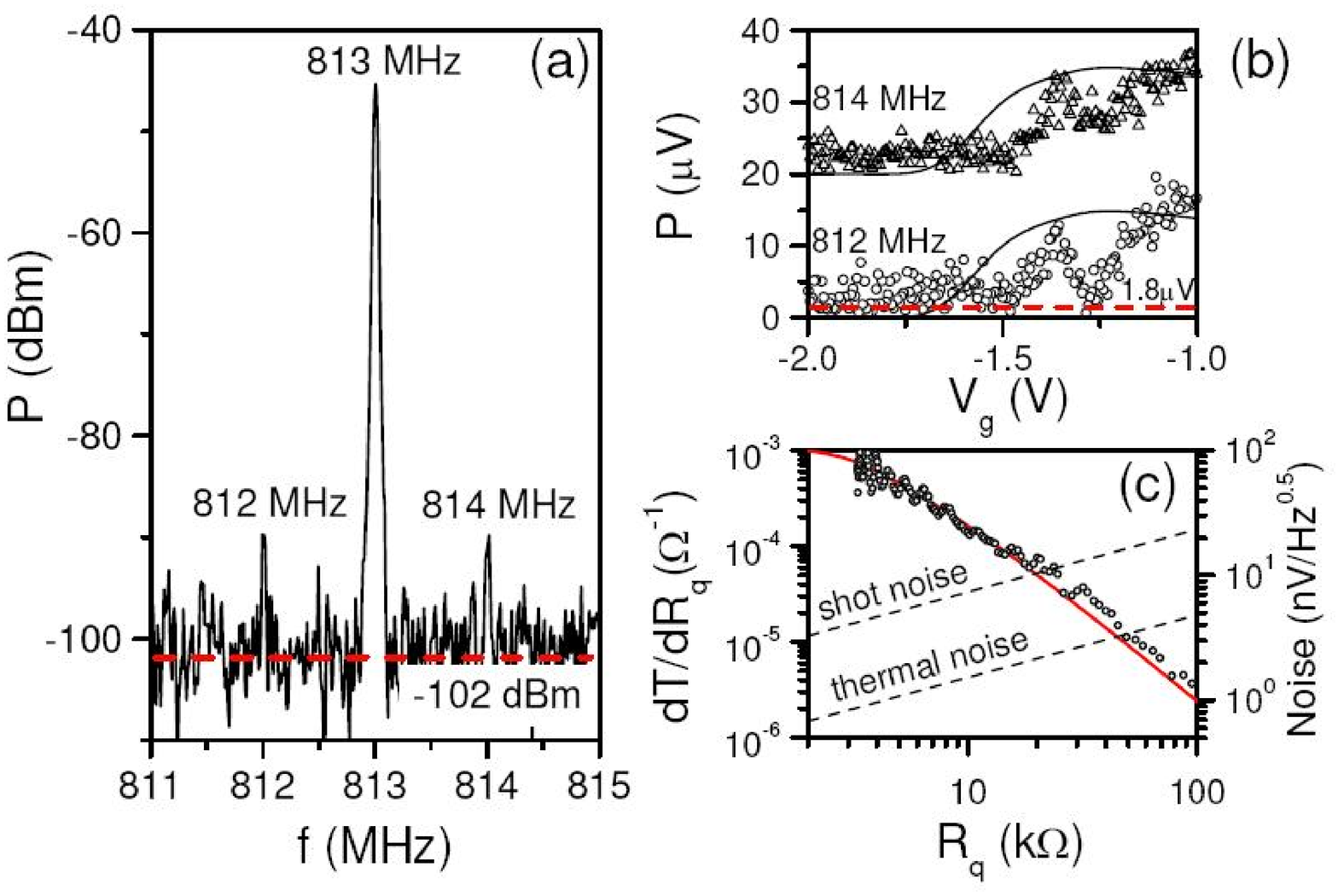}
\center{Qin {\it et al:~} Figure 3/3}
\end{figure}


\newpage
\begin{thebibliography}{10}
\bibitem{schoelkopf}
R. J. Schoelkopf, P. Wahlgren, A. A. Kozhevnikov, P. Delsing, and D. E. Prober, 
Science {\bf 280}, 1238 (1998). 

\bibitem{rimberg}
W. Lu, Z. Ji, L. Pfeiffer, K. W. West, and A. J. Rimberg, 
Nature (London) {\bf 423}, 422 (2003).

\bibitem{buehler}
T. M. Buehler, D. J. Reilly, R. P. Starrett, Andrew D. Greentree, 
A. R. Hamilton, A. S. Dzurak, and R. G. Clark, 
Appl. Phys. Lett. {\bf 86}, 143117 (2005). 

\bibitem{biercuk}
M. J. Biercuk, D. J. Reilly, T. M. Buehler, V. C. Chan, J. M. Chow, R. G. Clark, 
and C. M. Marcus, http://xxx.lanl.gov, cond-mat/0510550.

\bibitem{wood}
D. K. Wood, S. -H. Oh, S. -H. Lee, H. T. Soh, and A. N. Cleland, 
Appl. Phys. Lett. {\bf 87}, 184106 (2005). 

\bibitem{fujisawa}
T. Fujisawa and Y. Hirayama, Appl. Phys. Lett. {\bf 77}, 543 (2000).

\bibitem{field}
M. Field, C. G. Smith, M. Pepper, D. A. Ritchie, J. E. F. Frost, G. A. C. Jones, 
and D. G. Hasko, Phys. Rev. Lett. {\bf 70}, 1311 (1993). 

\bibitem{elzerman}
J. M. Elzerman, R. Hanson, J. S. Greidanus, L. H. Willems van Beveren, S. De Franceschi, 
L. M. K. Vandersypen, S. Tarucha, and L. P. Kouwenhoven
Phys. Rev. B {\bf 67}, 161308 (2003).

\bibitem{vandersypen}
L. M. K. Vandersypen, J. M. Elzerman, R. N. Schouten, L. H. Willems van Beveren, 
R. Hanson, and L. P. Kouwenhoven, Appl. Phys. Lett. {\bf 85}, 4394 (2004). 

\bibitem{cap}
The total capacitance of a PC device 
always includes a parasitic capacitance ($C_p$) 
due to the bond pads and nearby gate electrodes.
Comparing to $C_p$, the capacitance ($C_q$) 
from the point contact itself 
is usually much smaller: $C \approx C_p+C_q \approx C_p$. 

\bibitem{qin}
H. Qin, A. W. Holleitner, K. Eberl, and R. H. Blick,  
Phys. Rev. B {\bf 64}, 241302(R) (2001). 

\bibitem{noise}
In the experiment, the maximum RF voltage across the PC was $23~\mathrm{mV}$, 
resulting in a current $I_{ds} \approx 400~\mathrm{nA}$. 
Such a current significantly increases the shot noise to 
a level of $\sqrt{2eI_{ds}} \approx 0.4~\mathrm{pA/\sqrt{Hz}}$, 
equivalent to a noise voltage of about $0.4~\mathrm{\mu V}$ 
at the input port of the spectrum analyzer. 
At $4.2~\mathrm{K}$, the thermal noise ($\sqrt{4k_BT/R_q^0} \approx 80~\mathrm{fA/\sqrt{Hz}}$) 
seen by the spectrum analyzer is only $70~\mathrm{nV}$.

\bibitem{korotkov}
A. N. Korotkov and D. V. Averin, 
Phys. Rev. B. {\bf 64}, 165310 (2001).

\bibitem{jordan}
A. N. Jordan and M. B\"uttiker, 
Phys. Rev. Lett. {\bf 95}, 220401 (2005). 


\end{thebibliography}
\end{document}